\newcount\drawps\drawps=0 
 
\ifnum\drawps=1
\documentstyle[aps,twocolumn,prb]{revtex}
\else
\documentstyle[preprint,psfig,aps,prb]{revtex}
\fi
 
\input{epsf}
\begin{document}
 \draft
\title{
Spin Dynamics in the Two-Dimensional Spin 1/2 Heisenberg Antiferromagnet 
}
\author{
A.F. Albuquerque$^1$, A.\ S.\ T.\ Pires$^1$,
and
M.\ E.\ Gouv\^ea$^2$
}  
\date{April 20, 2005}
\address{
$^1$Departamento de F\'{\i}sica, ICEx,
Universidade Federal de Minas Gerais,
Belo Horizonte, \\CP 702, CEP 30123-970, MG, Brazil,\\
$^2$Centro Federal de Educa\c{c}\~ao Tecnol\'ogica de Minas Gerais,
Belo Horizonte,MG, Brazil
}
 
\maketitle
\ifnum\drawps=1
\widetext
\vskip-0.4in\hskip0.75in
\parbox[t]{5.5in}{
\else
\begin{abstract}
\fi
We present low-temperature dynamic properties of the quantum
two-dimensional antiferromagnetic Heisenberg model with spin $S=1/2$.
The calculation of the dynamic correlation function is performed
by combining a projection operator formalism
and  the  modified spin-wave theory (MSW), which gives a gap
in the dispersion relation for finite temperatures.
The so calculated dynamic correlation function shows a double peak
structure.
We also obtain the spin-wave damping and compare our results 
to some experimental data and theoretical results obtained
by other authors using different approaches.

\ifnum\drawps=1
 }
 \narrowtext
\else
 \end{abstract}
\vskip 2cm
\fi
      
\section{INTRODUCTION}
\label{Intro}  

Interest in quantum antiferromagnetism is old and can be traced
back to the early years of quantum theory, with Bethe's solution for
the Heisenberg antiferromagnetic chain \cite{Bethe}.
However, research in this field remains very active and was further triggered
by the discovery of high-temperature superconductivity in copper-oxide
compounds.
Superconductivity in the cuprates is attained upon doping the stoichiometric
parent compounds, such as $La_2CuO_4$, which are believed to be
experimental realizations of the two-dimensional quantum Heisenberg
antiferromagnet (2DQHAF) with spin $S=1/2$, described by the model Hamiltonian
\begin{equation}
\label{H1}
{\cal{H}} = J \sum_{\langle i,j \rangle}
\vec{S}_{\vec{r}_i} \cdot \vec{S}_{\vec{r}_j} ,
\end{equation}
where $J > 0$; ${\langle i,j \rangle}$ denotes the nearest neighbor (NN) sites
on a square lattice, without double counting of bonds.
 The 2DQHAF can be mapped into
the 2D quantum nonlinear $\sigma$ model\cite{Haldane} (2DQNLSM)
\cite{Haldane} and many theoretical works have been dedicated 
to the investigation of the properties of this last model.
However, the mapping is rigorously valid only in the large $S$ continuum limit
\cite{Affleck}, although it
can be justified on general grounds for the extreme
quantum limit \cite{Chak} $S=1/2$.

The widely held belief that antiferromagnetism plays a central role
in high-temperature superconductivity has contributed to
a noticeable proliferation of theoretical, numerical, and experimental
works devoted to the investigation of the magnetic properties of
the stoichiometric parent compounds, as described by (\ref{H1}) and by
the 2DNLSM \cite{Manousakis,Kastner}.
However, despite this connection with high-$T_c$ superconductivity,
the understanding of the properties of the system is important by itself.

Some early theoretical investigations of the 2DQHAF \cite{Anderson1}
raised doubts about the nature of the ground state of the model,
suggesting that it would be a disordered quantum spin-liquid state with 
correlations decaying exponentially with distance \cite{Anderson2}.
Later, further investigations ruled out this possibility and the system
is known to exhibit a broken symmetry N\'eel ground state 
\cite{Manousakis},
which is destroyed by thermal fluctuations \cite{Mermin}
when $T > 0$.
In fact, at low temperatures, the system is in a renormalized classical (RC)
regime, that is, it behaves as a classical 2D system with coupling constants
simply renormalized by quantum fluctuations, as showed by Chakravarty, 
Halperin and Nelson \cite{Chak} in their renormalization-group analysis 
of the 2DQNLSM.
This approach was improved by Hasenfratz 
and Niedermayer \cite{Hasen} who
obtained an expression for the correlation length $\xi$ of the 2DQNLSM in
the RC regime which agrees very well with experimental data for the monolayer
cuprate $Sr_2CuO_2Cl_2$ \cite{Greven1} 
over a certain temperature range.

At higher temperatures, $T/J \sim 0.5$, the 2DQNLSM shows a crossover
into a quantum critical (QC) region, with correlation length linear in $1/T$
\cite{Chubukov}.
Although this prediction is difficult to be experimentally tested in cuprates
such as $Sr_2CuO_2Cl_2$ and $La_2CuO_4$, given the high value of $J$
in these compounds ($J \sim 130meV$), recent experiments on copper formiate
tetradeuterate (CFTD) \cite{Ronnow1,Carreta}, 
which is also described as a
2DQHAF \cite{Ronnow2,Coldea} and whose much smaller 
value of $J$ ($J \sim 6meV$)
allows measurements up to higher temperatures, 
found no evidence for a crossover
into a QC phase.
It should be remarked that the 2DQNLSM is expected to model
the 2DQHAF in the limit of low temperatures, when the correlation length is
very large, and the above mentioned experiments seem to settle an upper 
temperature limit for the applicability of this approach.
In fact, the validity of the 2DQNLSM approach was shown
to be inadequate to describe the behavior of the 2DQHAF with spin value greater
than or equal to 1 in the experimentally acessible temperature
region \cite{Cuccoli2}.
For spin S=1/2, a series of experiments \cite{Greven1,Greven2} 
and Monte Carlo simulations 
\cite{Beard,Troyer,Elstner} showed that the length scale 
at which the renormalized 2DQNLSM description becomes valid
is surprisingly long.
The low energy spectrum of the S=1/2 Heisenberg model, obtained
by quantum Monte Carlo on finite size lattices\cite{Lavalle},
disagrees rather strongly with the prediction
of the non-linear sigma model even when the size of the system is not too
small.
All these results show that it is important to work 
directly with Hamiltonian (1)
instead of using the 2DQNLSM if
one wants to make comparisons with experimental data.

Nowadays, it is believed that the available experimental data for 
{\em{static}} properties is well described by a combination of 
low-temperature static properties \cite{Chak},
QMC simulations \cite{Beard,Troyer} at an intermediate
range, and, at higher temperatures,
by high-T expansion \cite{Elstner} and pure-quantum self-consistent
harmonic approximation (PQSCHA) \cite{Cuccoli}.
So, the attention has turned to the dynamical properties of the 2DQHAF at finite
temperatures, since the intrinsic non-linearities in the equations of motion
for a spin system give room for stronger quantum effects in the dynamics,
particularly at higher excitation energies.
Therefore, further microscopic calculations of the
dynamic structure factor for (\ref{H1}) are still very welcome in order
to allow a better understanding about the system's behavior at
finite temperatures.

Many of the interesting phenomena and experimental measurements in
strongly correlated quantum systems are related to the dynamics of the system.
Despite the considerable progress in many-body theory, available
exact results on quantum dynamics in many-body systems are
rather scarce.
Indeed, even a systematic framework for approximate calculations is not well
established \cite{Maeda}.

In this paper, we calculate the dynamical correlation function
for the 2DQHAF with $S=1/2$ using  the equation of motion 
approach in conjunction
with  projection operator methods, following a procedure
proposed originally by Reiter \cite{Reiter} and further
developed by other authors \cite{Raedt}.
This method has proven successful in the study of the classical
and quantum Heisenberg models in one \cite{Gouvea}
 and two \cite{Menezes}
dimensions showing good agreement with experimental data,
Molecular Dynamic simulations and, also, with other theories.

The calculation of the memory function, which plays a central
role in the formalism, does not require long-range order to be valid
because it depends only on correlations between nearest neighbors.
The frequency of the local spin-wave modes is also used as input in the method.
The calculation of the static correlations and spin-wave frequency
 needed in the formalism
we are using, is done within the context of the modified spin-wave (MSW)
theory proposed by Takahashi \cite{Takahashi} and
Hirsch and Tang \cite{Hirsch}.
The MSW theory applied to antiferromagnets imposes a constraint
on the total staggered magnetization to be zero in an
isotropic system, as required by Mermin-Wagner theorem \cite{Mermin}.
It has been shown \cite{Manousakis} that the results obtained 
via the MSW approximation are identical to the ones obtained
by Arovas and Auerbach \cite{Arovas} using a path integral
formulation in the Schwinger boson representation.
Also, the correlation length, as calculated within MSW theory,
agrees with one-loop order in renormalization-group theory
for the 2DQNL$\sigma$M \cite{Chak}.

The combination of these two techniques, the projection operator method and 
the MSW theory was already applied by some of us \cite{Pires3}
to study the low temperature properties of the quantum one-dimensional
Heisenberg model with spin $S=1$, where a gap is expected to occur.

The dynamical structure function  of (\ref{H1}) at finite temperature
was firstly calculated by Auerbach and Arovas \cite{Auerbach} as
the Fourier transform of the imaginary part of the
spin-density correlation function
\begin{equation}
S({\vec{q}},\omega,T) = \frac{1}{\pi}
\langle S^z({\vec{q}},\omega)S^z(-{\vec{q}},\omega) \rangle ,
\end{equation}
using the Schwinger boson mean-field representation.
The procedure includes processes where the incident particle
creates quasiparticle excitations as well as scattering from thermally
excited particles.
However, the projection operator method, in the approximation 
proposed by Reiter \cite{Reiter}, goes beyond a mean field theory 
and, thus, the effects  due to magnon scattering are more properly 
incorporated in the calculation of the dynamical structure
factor.
This  procedure was applied to the
2DQHAF model by Becher and Reiter \cite{Becher},
using a standard spin-wave formalism for the calculation of 
the static quantities,
however, as it is well known, 
conventional spin-wave theory predicts a zero gap
value for finite temperatures, i.e., it does not take into account the 
inexistence of long-range order (LRO) for $T > 0$. 
More recently, the spin dynamics of the 2DQHAF was investigated 
by Nagao and Igarashi \cite{Nagao} by using the self-consistent
theory of Blume and Hubbard \cite{Blume}.
Their method is expected to work well at high temperatures, 
although the authors claim that the results obtained 
for relatively low temperatures,
$ T/J \sim 0.4$, are reliable because the relaxation function
is found to satisfy a dynamical scaling relation consistent
with the nonlinear $\sigma$ model analysis and, also, Monte Carlo 
simulations \cite{Makivic}.

The outline of this paper is as follows:
Section (\ref{S2}) gives a brief overview of the MSW results
for the spin-wave energy, 
as obtained by Takahashi \cite{Takahashi}.
The steps leading to the calculation of the dynamical 
structure factor are also given in this section.
In Section (\ref{S3}), we discuss our numerical results 
and, finally, in Section (\ref{S4}) we present our conclusions.

\section{The Modified spin-wave theory and the projection operator method}
\label{S2}

It is well known that the standard spin-wave theory is not applicable to
low-dimensional quantum magnets at finite temperatures without
modifications \cite{Mermin}.
The consequence of the Mermin-Wagner theorem is enforced by hand in 
a variational density-matrix approach proposed by Takahashi \cite{Takahashi},
which we will shortly review here.
We start our calculations, writing (\ref{H1}) in the form
\begin{equation}
\label{Hamiltonian}
{\cal{H}} = J \sum_{\langle i,j \rangle}
\vec{S}_{\vec{r}_i} \cdot \vec{T}_{\vec{r}_j},
\end{equation}
where we divide the lattice into two sublattices $A$
and $B$: spins in sublattice $A$ ($B$) are denoted as $\vec{S}_{\vec{r}_i}$ 
($\vec{T}_{\vec{r}_j}$) and the sum runs over all ${\vec{r}_i} \in A$ sublattice
sites and its NN on the $B$ sublattice, avoiding double counting of bonds.
We now apply a Dyson-Maleev transformation to represent the spin operators 
in each antiferromagnetic sublattice in terms of bosonic operators
\begin{eqnarray}
\label{Dyson}
S^-_{\vec{r}_i}&=& a^{\dag}_{\vec{r}_i} ~~,~
S^{\dag}_{\vec{r}_i} = 
(2S - a^{\dag}_{\vec{r}_i} a_{\vec{r}_i}) a_{\vec{r}_i}~~,
\nonumber \\
S^z_{\vec{r}_i} &=& S - a^{\dag}_{\vec{r}_i} a_{\vec{r}_i}~~, \nonumber \\
T^-_{\vec{r}_j} &=& -b_{\vec{r}_j}~~,~~ 
T^{\dag}_{\vec{r}_j} = - b^{\dag}_{\vec{r}_j} 
(2S - b^{\dag}_{\vec{r}_j}b_{\vec{r}_j})~~,
 \\
T^z_{\vec{r}_j} &=& -S + b^{\dag}_{\vec{r}_j} b_{\vec{r}_j}~~,
\end{eqnarray}
following the canonical commutation relations.
The Hamiltonian (\ref{Hamiltonian}) becomes,
\begin{eqnarray} 
\label{Hab}
{\cal{H}} &=& -2NJS^2 + J \sum_{\langle i,j \rangle} \left\{ 
S \left( a^{\dag}_{\vec{r}_i} a_{\vec{r}_i} + b^{\dag}_{\vec{r}_j} b_{\vec{r}_j}
- a_{\vec{r}_i} b_{\vec{r}_j} - a^{\dag}_{\vec{r}_i} b^{\dag}_{\vec{r}_j} \right)
+ \frac{1}{2} a^{\dag}_{\vec{r}_i} (a_{\vec{r}_i} - 
b^{\dag}_{\vec{r}_j})^2 b_{\vec{r}_j}
\right\}~~.
\end{eqnarray}
We then introduce an ideal spin-wave ansatz for the density matrix of the
system
\begin{equation}
\label{denmatrix}
\rho = \exp \left\{ \frac{1}{T} \sum_{\vec{q}} ~'~ \omega_{\vec{q}} 
(\alpha^{\dag}_{\vec{q}} \alpha_{\vec{q}} +
 \beta^{\dag}_{-\vec{q}} \beta_{-\vec{q}}) \right\}.
\end{equation}
$\sum_{\vec{q}}^{'}$ indicates summation over half of the first Brillouin zone.
$\alpha_{\vec{q}}$ and $\beta^{\dag}_{-\vec{q}}$ are given by the Bogoliubov
transformation
\begin{eqnarray} 
\label{Bogoliubov}
\alpha_{\vec{q}} &=& \cosh (\theta_{\vec{q}}) a_{\vec{q}} - 
\sinh (\theta_{\vec{q}}) b^{\dag}_{-\vec{q}} ~,
\nonumber \\
\beta^{\dag}_{-\vec{q}} &=& - \sinh (\theta_{\vec{q}}) a_{\vec{q}} + 
\cosh (\theta_{\vec{q}}) b^{\dag}_{-\vec{q}},
\end{eqnarray}
where we introduced the Fourier transform of the original boson operators,
\begin{eqnarray}
\label{Fourier}
a_{\vec{q}} &=& 
\frac{1}{\sqrt{N_A}} \sum_{\vec{r}_i \in A} 
e^{-i\vec{q}\cdot\vec{r}_i} a_{\vec{r}_i},
\nonumber \\
b^{\dag}_{-\vec{q}} &=& \frac{1}{\sqrt{N_B}} \sum_{\vec{r}_j \in B} 
e^{-i \vec{q} \cdot \vec{r}_j} b^{\dag}_{\vec{r}_j} .
\end{eqnarray}
$N_A + N_B = N $ is the number of sites in the lattice.
The dispersion relation, $\omega_{\vec{q}}$, is obtained after minimizing
the free energy with the constraint that the magnetization on
each sublattice is zero, $\langle S^z_{\vec{r}} \rangle = 0$ and
$\langle T^z_{\vec{r}} \rangle = 0$, as required by Mermin-Wagner theorem.
In this way, we get
\begin{eqnarray}
\label{dispersion}
\omega_{\vec{q}} &=& \lambda ( 1 - \eta^2 \gamma^{2}_{\vec{q}} )^{1/2} ,
\\
\gamma_{\vec{q}} &=& \frac{1}{2} \left( cos q_x + cos q_y \right)  .
\nonumber 
\end{eqnarray} 
We will write the wavevector $\vec{q} = (q_x,q_y)$ 
in units of the inverse lattice spacing.

The parameters $\eta$ and $\lambda$ can be determined by solving the following
set of self-consistent equations,
\begin{eqnarray}
\label{selfconsistent}
S + \frac{1}{2} &=& \frac{2}{N}
\sum_{\vec{q}} ~'~ \frac{1}{(1 - \eta^2 \gamma^{2}_{\vec{q}})^{1/2}}
\frac{1}{2}
\coth \left\{
\frac{\lambda}{2T} (1 - \eta^2 \gamma^{2}_{\vec{q}})^{1/2}
\right\}, \nonumber
\\
\frac{\eta \lambda}{4 J} &=& \frac{2}{N}
\sum_{\vec{q}} ~'~ \frac{\eta \gamma^{2}_{\vec{q}}}
{(1 - \eta^2 \gamma^{2}_{\vec{q}})^{1/2}}
\frac{1}{2}
\coth 
\left\{
\frac{\lambda}{2T} (1 - \eta^2 \gamma^{2}_{\vec{q}})^{1/2}
\right\}.
\end{eqnarray}
The same equations have been obtained by Hirsch and Tang \cite{Hirsch},
and, as said before, by Arovas and Auerbach \cite{Arovas} in their Schwinger
boson treatment.

Takahashi \cite{Takahashi} was able to find out the asymptotic forms
of (\ref{selfconsistent}) in the $T\rightarrow 0 $ limit and, also,
to evaluate the  $\eta$ parameter for the spin $S=1/2$ case
for $4 \times 4$ and $64 \times 64$ lattices.
However, for the  calculation of the dynamical structure factor
according to the projection operator  procedure, we need to know the 
spin-wave energy for the infinite square lattice model at finite
temperatures and, thus, we solved Eqs. (\ref{selfconsistent}) using 
an iterative numerical model obtaining the results displayed in Table I.

As it is well known, there is no gap in the 2DQHAF with $S=1/2$ at $T=0$.
Indeed, when $T\rightarrow 0 $, we can see that $\eta \rightarrow 1 $.
However, for finite temperatures, $\eta$ becomes smaller than unity and so
a gap opens in the system, reflecting the fact that 
the correlation length, $\xi$,
becomes finite and the long wavelength, low-energy spin-waves cannot
propagate. 
So, spin-wave excitations are well defined only for wavelengths significantly
smaller than $\xi$.

Now, we describe only the main steps leading to the calculation of
the dynamic structure factor following the projection operator method.
A complete description of the theory can be found in Refs.
\onlinecite{Reiter,Raedt}.
One of the advantages of this procedure is that it allows us to obtain the 
structure factor for all values of the wavevector ${\vec{q}}$, 
while calculations
based on the nonlinear $\sigma$ model are restricted to the long wavelength
limit ${\vec{q}} \rightarrow (0,0)$.

 In antiferromagnets, spin-waves have two flavors, one associated with
the conventional magnetization, ${\cal{M}}^{\alpha}_{\vec{q}} = S^{\alpha}_{\vec{q}} 
+ T^{\alpha}_{\vec{q}}$, and another to the staggered magnetization,
${\cal{R}}^{\alpha}_{\vec{q}}= S^{\alpha}_{\vec{q}} - T^{\alpha}_{\vec{q}}$,
with $\alpha = x,y,z$.
Since the last is a lower energy mode, we will concentrate on the calculation
of the dynamical response associated to it.
We can see that the staggered magnetization is linear in magnon
creation and annihilation operators while the uniform
magnetization is a two-magnon operator process. 
In fact, the calculation of the ${\cal{M}}_{\vec{q}}$ correlation function
could be performed by
using the same procedure applied in this work, but the calculation would
require us to go to a higher order in magnon operators.

It is important to emphasize that
 we only calculate rotationally invariant quantities such as
${\cal{R}}_{\vec{q}} = \frac{1}{3} ( {\cal{R}}^{x}_{\vec{q}} + 
{\cal{R}}^{y}_{\vec{q}} + {\cal{R}}^{y}_{\vec{q}} )$.
The Fourier transform of the relaxation function is given by \cite{Pires}
\begin{equation}
\label{strucfac}
R ({\vec{q}},\omega) =
\frac{1}{2 \pi} \int_{-\infty}^{+\infty}
d t e^{- i \omega t}
\frac{ \left( {\cal{R}}_{\vec{q}} (t), {\cal{R}}_{\vec{q}} (0) \right) }
{ \left({\cal{R}}_{\vec{q}} , {\cal{R}}_{\vec{q}} \right) }~.
\end{equation}
Here, $(A,B)$ is the Kubo inner product of two operators $A$ and $B$ defined as
\cite{Mori}
\begin{equation}
\label{Kubo}
\left( A , B \right) =
\frac{1}{\beta}
\int_0^{\beta}
\langle e^{\lambda H} A^{\dag} e^{-\lambda H} B \rangle d \lambda ,
\end{equation}
where $<.~.~.>$ denotes the usual thermal average and $\beta = 1/ k_BT$. 
One can show that, after some analytical work, the dynamical correlation 
function $R({\vec{q}},\omega)$ is given by
\begin{equation}
\label{R1}
R ({\vec{q}}, \omega) =
\left({\cal{R}}_{\vec{q}} , {\cal{R}}_{\vec{q}} \right)
\langle \omega^2_{\vec{q}}\rangle
\frac{\Sigma_{\vec{q}}^{''} (\omega)}
{ \left[
 \omega^2 - \langle \omega^2_{\vec{q}} \rangle
+ \omega \Sigma^{'}_{\vec{q}} (\omega) \right]^2
 + \left[ \omega \Sigma{''}_{\vec{q}} (\omega) \right]^2
 }
\end{equation}
where $\Sigma^{'}_{\vec{q}}(\omega)$ and $\Sigma^{''}_{\vec{q}}(\omega)$
are the real and imaginary parts of the second order
memory function, $\Sigma_{\vec{q}}(\omega)$, respectively.
In time space, this memory function is expressed by
\begin{equation}
\label{memo}
\Sigma_{\vec{q}} (t) = \frac{\left(
 Q L^2 {\cal{R}}_{\vec{q}}, e^{-i QLQt} QL^2 {\cal{R}}_{\vec{q}}
\right)}
{\left( L {\cal{R}}_{\vec{q}}, L {\cal{R}}_{\vec{q}} \right)}   
\end{equation}
where $Q$ is a projection operator that projects out any term
proportional to ${\cal{R}}_{\vec{q}}$ and $ L {\cal{R}}_{\vec{q}}$, and
$L$ is the Liouville operator, defined by the relation $LA=-i [A,H]=-i \dot{A}$.
Reiter \cite{Reiter} has shown that, to leading order in temperature,
 the projection operator $Q$ in the exponential function in 
(\ref{memo}) can be dropped and we can also write
\begin{equation}
\label{Qoper}
QL^2  {\cal{R}}_{\vec{q}} = L^2 {\cal{R}}_{\vec{q}} -
\langle \omega^2_{\vec{q}} \rangle {\cal{R}}_{\vec{q}}~~~.
\end{equation}
In (\ref{R1}), we also need to define the second frequency moment
$\langle \omega^2_{\vec{q}}\rangle$, which is given by
\begin{equation}
\label{secmom}
\langle \omega^2_{\vec{q}} \rangle =
\frac{\left( L {\cal{R}}_{\vec{q}}, L {\cal{R}}_{\vec{q}} \right)}
{ \left( {\cal{R}}_{\vec{q}}, {\cal{R}}_{\vec{q}} \right) } .
\end{equation}

The second time derivatives needed to evaluate the numerator of the
memory function (\ref{memo}) are directly obtained from the definition of
the Liouville operator and from (\ref{Hamiltonian}).
Since this calculation is very straightfoward and the expressions are
enormous, we will not show them here.
We then apply the Dyson-Maleev (\ref{Dyson}) and Bogoliubov (\ref{Bogoliubov})
transformations for the spin operators in those expressions.
Doing so, we can simply replace the time evolution $\exp (-i L t)$ by the
harmonic time evolution 
\begin{eqnarray}
\label{tevol}
\alpha_{\vec{q}} (t) = e^{-i \omega_{\vec{q}} t} \alpha_{\vec{q}} (0)
\\
\alpha^{\dag}_{\vec{q}} (t) = e^{i \omega_{\vec{q}} t} \alpha^{\dag}_{\vec{q}} (0) ,
\end{eqnarray}
with similar equations for $\beta_{\vec{q}} (t)$ and $\beta^{\dag}_{\vec{q}} (t)$.
So, we are left with a number of Kubo products of four bosonic operators which
can be decoupled by means of Wick's theorem.
After a tedious but simple calculation, we get
\begin{equation}
\label{memo2}
\Sigma_{\vec{q}} (t) = \frac{4}{N} \sum_{\vec{p}} ~'~ \left\{
A_{+} (\vec{q},\vec{p}) \cos (\Omega_{+} t) +
A_{-} (\vec{q},\vec{p}) \cos (\Omega_{-} t)  \right\} .
\end{equation}
$A_{+} (\vec{q},\vec{p})$ and $A_{-} (\vec{q},\vec{p})$ are given by
\begin{eqnarray}
\label{Apm}
A_{+} (\vec{q},\vec{p}) &=& \frac{\lambda^2 T n_{\vec{q}_{+}} n_{\vec{q}_{-}} 
( e^{\beta \Omega_{+}} - 1 ) }
{4 \Omega_{+} \omega_{\vec{q}_{+}} \omega_{\vec{q}_{-}} 
\left( L {\cal{R}}_{\vec{q}}, L {\cal{R}}_{\vec{q}} \right)}
\left[ s (\vec{q},\vec{p}) - t (\vec{q},\vec{p})  \right]^2 ,
 \\
A_{-} (\vec{q},\vec{p}) &=& \frac{\lambda^2  T n_{\vec{q}_{+}} n_{\vec{q}_{-}}
( e^{\beta \omega_{\vec{q}_{+}}} - e^{\beta \omega_{\vec{q}_{-}}} ) }
{4 \Omega_{-}  \omega_{\vec{q}_{+}} \omega_{\vec{q}_{-}} 
\left( L {\cal{R}}_{\vec{q}}, L {\cal{R}}_{\vec{q}} \right)}
\left[ s (\vec{q},\vec{p}) + t (\vec{q},\vec{p})  \right]^2  ,
\end{eqnarray}
where we introduced the notation 
$\vec{q}_{\pm} = \frac{\vec{q}}{2} \pm \vec{p}$,
and
\begin{eqnarray}
\label{st}
s (\vec{q},\vec{p}) &=&
 \left[ 4 J^2 (\gamma_{\vec{q_{+}}} + \gamma_{\vec{q_{-}}})
(\gamma_{\vec{q_{+}}} + \gamma_{\vec{q_{-}}} - 2) + \langle \omega^2_{\vec{q}} 
\rangle \right]
(1 + \eta \gamma_{\vec{q_{+}}})^{1/2} (1 + \eta \gamma_{\vec{q_{-}}})^{1/2} ,
\nonumber
\\
t (\vec{q},\vec{p}) &=& \left[ 4 J^2 (\gamma_{\vec{q_{+}}} + \gamma_{\vec{q_{-}}})
(\gamma_{\vec{q_{+}}} + \gamma_{\vec{q_{-}}} + 2) + \langle \omega^2_{\vec{q}} 
\rangle \right]
(1 - \eta \gamma_{\vec{q_{+}}})^{1/2} (1 - \eta \gamma_{\vec{q_{-}}})^{1/2} .
\end{eqnarray}
In the expressions above, 
$n_{\vec{q}} = (\exp (\beta \omega_{\vec{q}}) - 1)^{-1}$
is the boson occupation number and 
$\Omega_{\pm} (\vec{q},\vec{p})$ is defined as
\begin{equation}
\label{omepm}
\Omega_{\pm} (\vec{q},\vec{p})= \omega_{\vec{q}_{+}} \pm \omega_{\vec{q}_{-}}.
\end{equation}

The second moment is readily evaluated from its definition (\ref{secmom}).
It is given by the ratio of the following expressions
\begin{equation}
\label{numer}
\left( L {\cal{R}}_{\vec{q}}, L {\cal{R}}_{\vec{q}} \right) = 
\frac{4 J T}{N}
\sum_{\vec{k}} ~'~ \left( 2 \gamma_{\vec{k}} + \gamma_{\vec{k}+\vec{q}} +
\gamma_{\vec{k}-\vec{q}} \right)
\frac{\eta \gamma_{\vec{k}}}{(1 - \eta^2 \gamma^{2}_{\vec{q}})^{1/2}}
\coth \left\{
\frac{\lambda}{2T} (1 - \eta^2 \gamma^{2}_{\vec{q}})^{1/2}
\right\} ,
\end{equation}
and
\begin{equation}
\label{denomi}
\left( {\cal{R}}_{\vec{q}}, {\cal{R}}_{\vec{q}} \right) = 
\frac{T}{\lambda} \frac{1}{\left( 1 - \eta \gamma_{\vec{q}} \right)} .
\end{equation}
We note that (\ref{numer}) is also needed in the evaluation of equation
(\ref{Apm}).

If we take the Laplace transform of the memory function (\ref{memo2}) and then
apply the Cauchy formula we finally get the real and imaginary parts of the 
memory function as given by
\begin{eqnarray}
\label{real}
\Sigma^{'}_{\vec{q}} (\omega) = \frac{1}{2 \pi^{2}} {\cal{P}} \left\{
\int d^{2} p [ \frac{A_{+} (\vec{q},\vec{p})}{\omega + \Omega_{+} (\vec{q},\vec{p})}
+ \frac{A_{+} (\vec{q},\vec{p})}{\omega - \Omega_{+} (\vec{q},\vec{p})} +
\frac{A_{-} (\vec{q},\vec{p})}{\omega + \Omega_{-} (\vec{q},\vec{p})} +
\frac{A_{-} (\vec{q},\vec{p})}{\omega - \Omega_{-} 
(\vec{q},\vec{p})} ] \right\},
\end{eqnarray}
and
\begin{eqnarray}
\label{imag}
\Sigma^{''}_{\vec{q}} (\omega) = \frac{1}{2 \pi} 
\int d^{2} p [ A_{+} (\vec{q},\vec{p}) \delta(\omega + \Omega_{+} (\vec{q},\vec{p})) +
A_{+} (\vec{q},\vec{p}) \delta(\omega - \Omega_{+} (\vec{q},\vec{p})) +
\nonumber
\\
A_{-} (\vec{q},\vec{p}) \delta(\omega + \Omega_{-} (\vec{q},\vec{p})) +
A_{-} (\vec{q},\vec{p}) \delta(\omega - \Omega_{-} (\vec{q},\vec{p})) ].
\end{eqnarray}

Comparing equation (\ref{memo2}) with the corresponding one obtained by Becher
and Reiter (see equation (9) in their paper \cite{Becher}), we can note
that they are very similar if we assign to $\eta$ and $\lambda$ their zero
temperature values in our expressions.
But there are some slight differences between our definitions
 for $s (\vec{q},\vec{p})$
and $t (\vec{q},\vec{p})$, eqs. (\ref{st}), and their equivalent ones
due to the fact that, in this work, we considered two sub-lattices 
for the antiferromagnet. 

In the next section we discuss our numerical results for the dynamic
structure factor calculated from the above expressions.

\section{Discussion of the numerical results}
\label{S3}

The evaluation of the imaginary part of the memory function is
far from trivial.
For this purpose, we adopt a numerical method introduced by Gilat
and co-workers \cite{Gilat} and further developed by 
Wysin \cite{Gary1}.
A typical result is shown in figure 1, for $(q_x,q_y)=(\pi/2,\pi/2)$, 
$\tau=T/J=0.20$;
the frequency is given in units of $J S^2$ (throughout this paper, $k_{B}=\hbar=1$).
A prominent feature displayed in that figure 
is that $\Sigma^{''}_{\vec{q}} (\omega)$ vanishes
in a finite interval approximately centred at the zero 
temperature spin-wave energy.
This can be understood if we look at eq. (\ref{imag}).
We can see that we only have contributions when the argument of the delta functions,
given by differences or sums of the frequency and the functions
$\Omega_{-} (\vec{q},\vec{p})$ (which accounts for processes involving an
absorption followed by the reemission of a magnon with a lower energy) and
$\Omega_{+} (\vec{q},\vec{p})$ (that describes the decay of a magnon into
two others), vanishes.
The shape of these two functions is shown in figure 2, also for $(q_x,q_y)=(\pi/2,\pi/2)$ and
$\tau=0.20$.
We note that the non-unitary value of $\eta$ at finite temperatures, a
consequence of the absence of long-range order, imposes the opening of a small gap
between $\Omega_{-} (\vec{q},\vec{p})$ and $\Omega_{+} (\vec{q},\vec{p})$, as
we can see in the figure.
This gap is close to the zero-temperature spin-wave frequency.
So, as we raise the frequency, the two terms involving $\Omega_{-} (\vec{q},\vec{p})$
cease to contribute to the integral in a region where the term in
$\delta(\omega -\Omega_{+} (\vec{q},\vec{p}))$ still does not contribute,
and $\Sigma^{''}_{\vec{q}} (\omega)$ is zero within it.
A similar behavior was obtained by some of us in a previous work regarding
the dynamics of the $S=1$ antiferromagnetic chain \cite{Pires3}.
%


The real part of the memory function is readily evaluated by a generalization 
\cite{Gary2} of a numerical method used in one-dimensional cases.
In this way, we can directly compute the dynamical correlation function
by means of eq. (\ref{R1}).
In figure 3, we show the results obtained for
$(q_x,q_y)=(\pi/2,\pi/2)$ at $\tau=0.05$, $0.10$, $0.15$ and $0.20$.
The results for other values of $\vec{q}$ are very similar, except
for small wave-vectors
(such as $(q_x,q_y)=(\pi/64,\pi/64)$) at the highest temperatures studied.
In this case, we observe that the imaginary part of the memory function
changes its shape, from the {\it normal} one as displayed in figure 1 at
the lowest temperature investigated ($\tau=0.05$), into an anomalous one
in higher temperatures.
This leads to a {\it blurrying} of the peaks observed in $R(q,\omega)$.
However, this region -- small wavevector and high temperature --
is outside the validity range of the method employed in this work.

We can note that the cancellation of the imaginary part of the memory
function in an interval, as discussed above, leads to the vanishing
of the dynamic structure factor in the same region.
In particular, this feature prevents us to investigate 
the ${\vec{q}} \rightarrow (0,0)$
limit, since the gap remains relatively large and 
there are no peaks in $R(q,\omega)$.
So, we cannot compare our results to those obtained by Becher and Reiter
\cite{Becher}, who found that the damping of the magnons is zero at $T=0$.

But, what is remarkable in the curves obtained for $R(q,\omega)$
is the presence of a double-peak
structure.
It is interesting to note that the low-energy peak follows 
the behavior expected
for a spin-wave peak, that is, it becomes broader as the temperature increases
and its intensity decreases.
However, the high-energy peak surprisingly has its intensity {\it increased}
when we turn up the temperature.
One can be tempted to relate this peak to two-magnons excitations, which should
become more important when the temperature raises.
It is hard to imagine that such a sharp structure defining a double-peak
structure will be destroyed
if we take into account higher order processes, which are expected to
give important contributions only at high temperatures.
A similar structure was also 
obtained by Auerbach and Arovas
\cite{Auerbach} in their Schwinger boson treatment of (\ref{H1}).
Although we cannot assure that such a result, a double peak structure,
is not an artifact of the approximations done, it should
be worth to check if such a
structure could be resolved in neutron scattering experiments.
The same suggestion was made by Auerbach and Arovas.

In order to calculate the damping of the spin-waves, $\Gamma (\vec{q},\tau)$,
we fit lorentzians to the data points obtained for $R ({\vec{q}}, \omega)$.
$\Gamma (\vec{q},\tau)$ is so obtained as the half width of the
low-energy peak.
In figure 4, we show $\Gamma (\vec{q},\tau)$ as a function of the
wave-vetor magnitude along two high symmetry directions 
in the antiferromagnetic Brillouin zone, for $\tau=0.05$.
We then choose the wave-vector $(q_x,q_y)=(\pi/4,\pi/4)$, indicated
in the figure, since this has an intermediate value for the damping, in order
to compare the temperature dependence of $\Gamma (\vec{q},\tau)$ with
the experimental data obained by Thurber {\it et al.} \cite{Thurber}.
We show the results in figure 5.
We can see an excelent agreement between our data, indicated by filled diamonds,
and the experimental ones, up to a temperature of $T \sim 350K$.
Also shown in the figure are the previous results by Ty$\check{c}$ and
Halperin \cite{Tyc} in their self-consistent calculations and we also
see a good agreement between our results.

\section{Conclusions}
\label{S4}

In summary, we have calculated the dynamic structure factor 
for the antiferromagnetic
Heisenberg model with $S=1/2$ using a projection operator technique.
This approach was also followed by Becher and Reiter \cite{Becher},
but they used conventional spin-wave theory in order to calculate the static
correlations needed as an input.
Instead, we combined the method with the MSW theory,
which goes beyond the linear one in the sense that it takes into account
the inexistence of long-range order in the model at any finite temperature.
We obtained a double-peak structure, as in a previous work
of Auerbach and Arovas \cite{Auerbach}.
The damping, calculated by fitting lorentzians to our data points, agrees
well with experimental data \cite{Thurber} and with previous theoretical
calculations by Ty$\check{c}$ and Halperin \cite{Tyc} up to a temperature
$T \sim 350K$.

It is worth to remark that, in the classical limit,
the gap will vanish and the double peak struture will
disappear. 
So, the double peak structure found here is a quantum effect.

Huberman et al \cite{Huberman} probed the low temperature magnetic excitations 
of the 2D S=5/2 AF compound $Rb_2MnF_4$ using pulsed
inelastic neutron scattering and found a dominant sharp peak 
that can be identified with one-magnon excitations.
However, in addition to this one magnon peak, he was able to 
observe a relatively weak continuum scattering at higher energies.
This continuum scattering was attributed to
scattering by pairs of magnons as expected to happen for
the ${\cal{M}}_q$ correlation function. 
This will be the  subject of a future work.

\vskip 1cm
{\small{Acknowledgements: We acknowledge Takashi Imai 
for sharing the experimental data
shown in figure 5. 
This work was partially  supported by CNPq (Conselho Nacional para o
Desenvolvimento Cient\'{\i}fico e Tecnol\'ogico)-Brazil}.}

\begin{table}
\begin{center}
\label{Table1}
\begin{tabular}{|c|c|c|}
\hline
 T/J & $\eta$ & $\lambda$ \cr
 \hline
 0.05 &  0.999999993  &  2.609963279 \cr
 0.10 &  0.999995178  &  2.609876929 \cr
 0.15 &  0.999939299  &  2.599595679 \cr
 0.20 &  0.999709621  &  2.562049716 \cr
 0.25 &  0.999280433  &  2.567211935 \cr
 0.30 &  0.998523449  &  2.558794763 \cr
\hline
\end{tabular}
\end{center}
\caption{Results for the $\eta$ and $\lambda$ parameters for some temperatures
$T$ used in our calculations. The values were obtained by solving Eqs (11)
numerically.}
\end{table}
\begin{figure}
\begin{center}
\label{fig1}
\epsfxsize=16cm
\epsfysize=13cm
\epsfbox{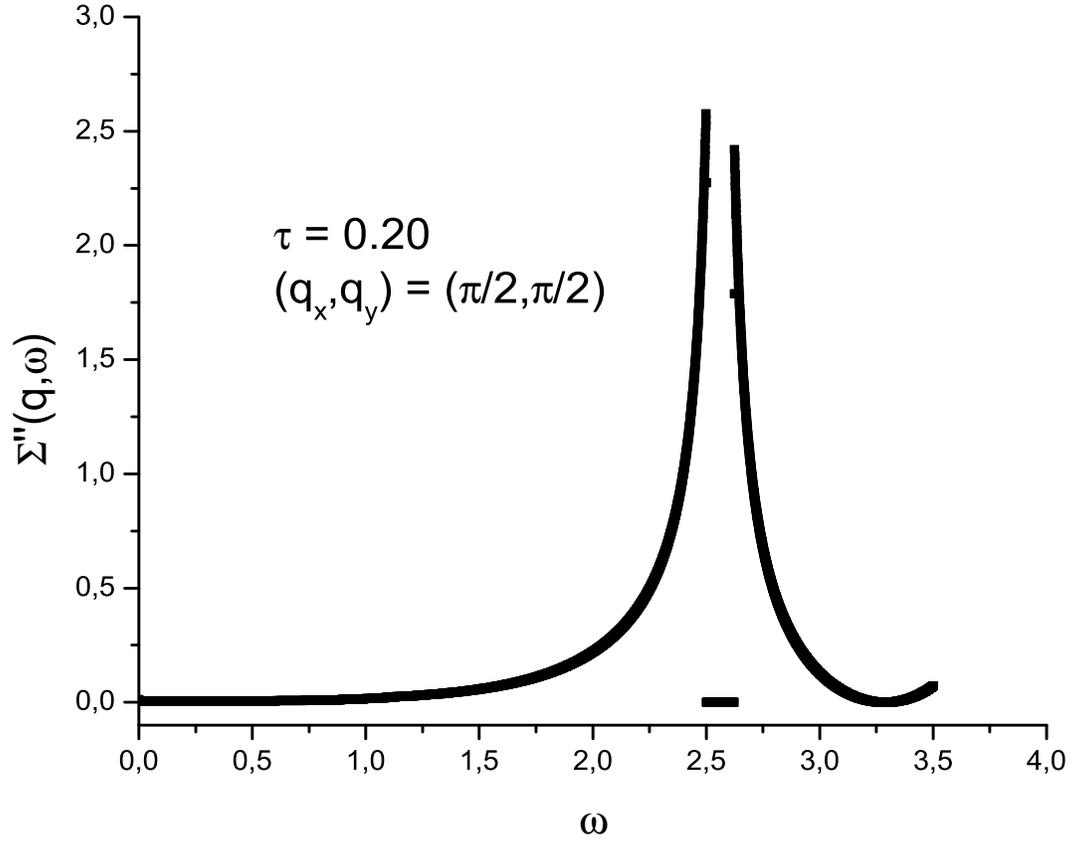}
\end{center}
\caption{Imaginary part of the memory function, 
$\Sigma^{''}_{\vec{q}} (\omega)$, as
a function of $\omega$ for $(q_x,q_y)=(\pi/2,\pi/2)$ and $\tau=0.20$.} 
\end{figure}
\begin{figure}
\begin{center}
\label{fig2}
\epsfxsize=16cm
\epsfysize=13cm
\epsfbox{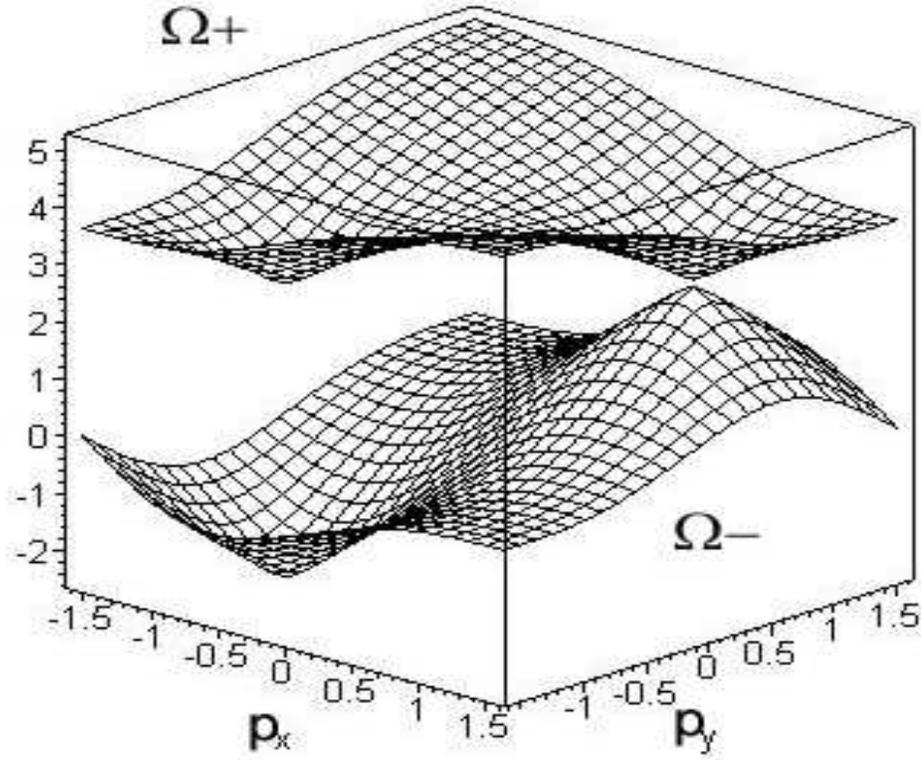}
\end{center}
\caption{$\Omega_{-} (\vec{q},\vec{p})$ and $\Omega_{+} (\vec{q},\vec{p})$ as
a function of $(p_x,p_y)$ for $(q_x,q_y)=(\pi/2,\pi/2)$ and $\tau=0.20$.} 
\end{figure}
\begin{figure}
\begin{center}
\label{fig3}
\epsfxsize=16cm
\epsfysize=13cm
\epsfbox{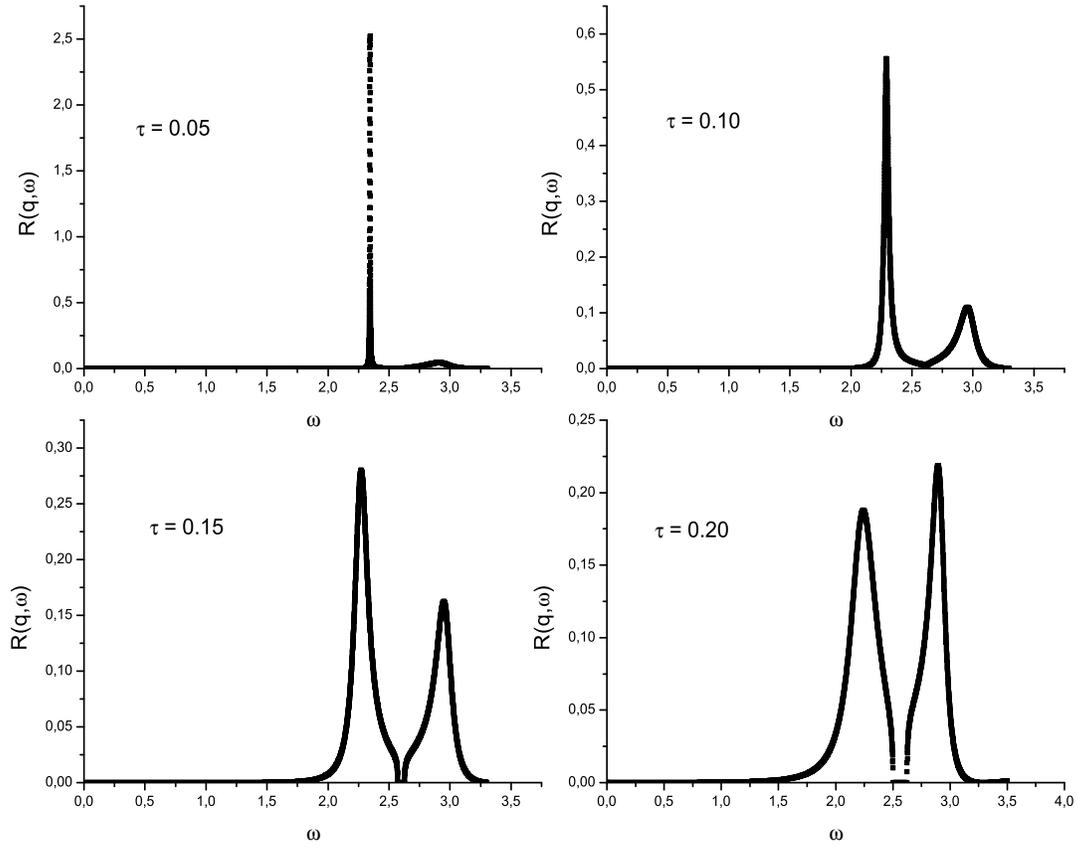}
\end{center}
\caption{Dynamic structure factor, 
$R(q,\omega)$, for $(q_x,q_y)=(\pi/2,\pi/2)$ and
$\tau=0.05$,$0.10$,$0.15$ and $0.20$.} 
\end{figure}
\begin{figure}
\begin{center}
\label{fig4}
\epsfxsize=16cm
\epsfysize=13cm
\epsfbox{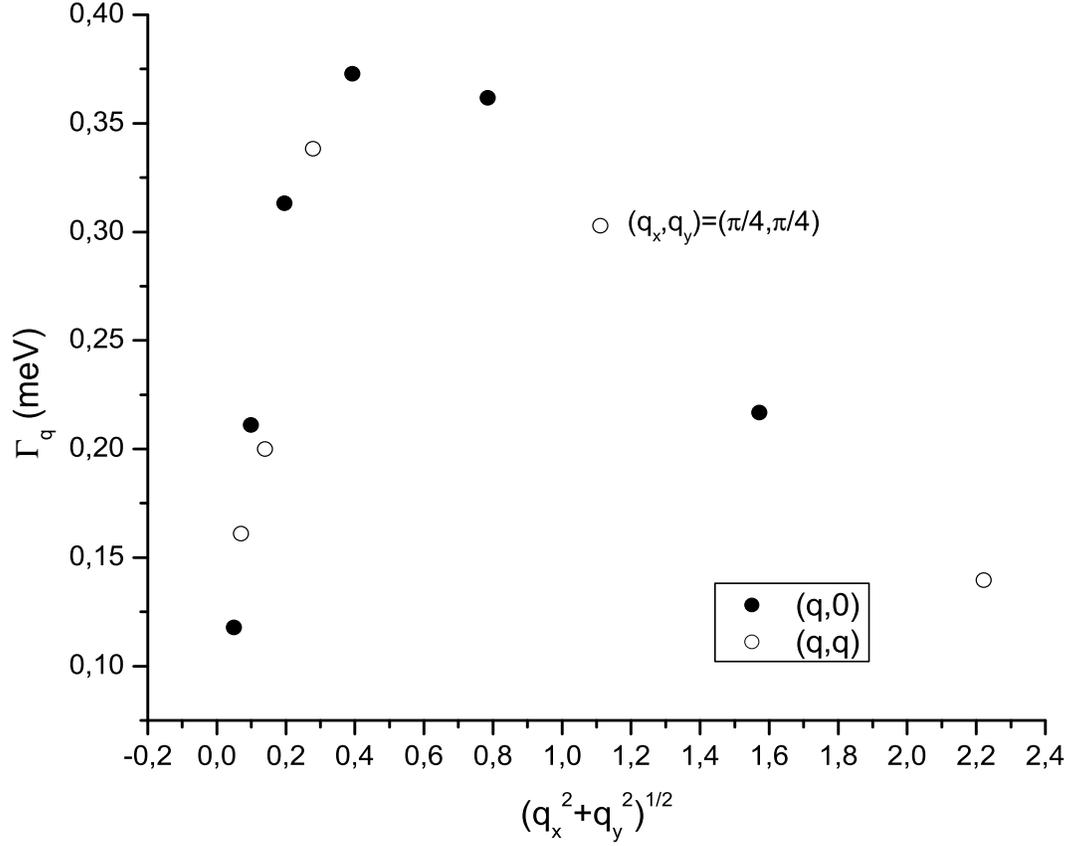}
\end{center}
\caption{
Results obtained in our work (as explained in the text) for the
spin-wave damping, $\Gamma (\vec{q},\tau)$,as a function of
$(q_{x}^{2}+q_{y}^{2})^{1/2}$ for $\tau=0.05$. Filled circles are for wavectors along the
direction $(q,0)$ and open circles for $(q,q)$. 
Wave-vector $(q_x,q_y)=(\pi/4,\pi/4)$
is indicated. } 
\end{figure}
\begin{figure}
\begin{center}
\label{fig5}
\epsfxsize=16cm
\epsfysize=13cm
\epsfbox{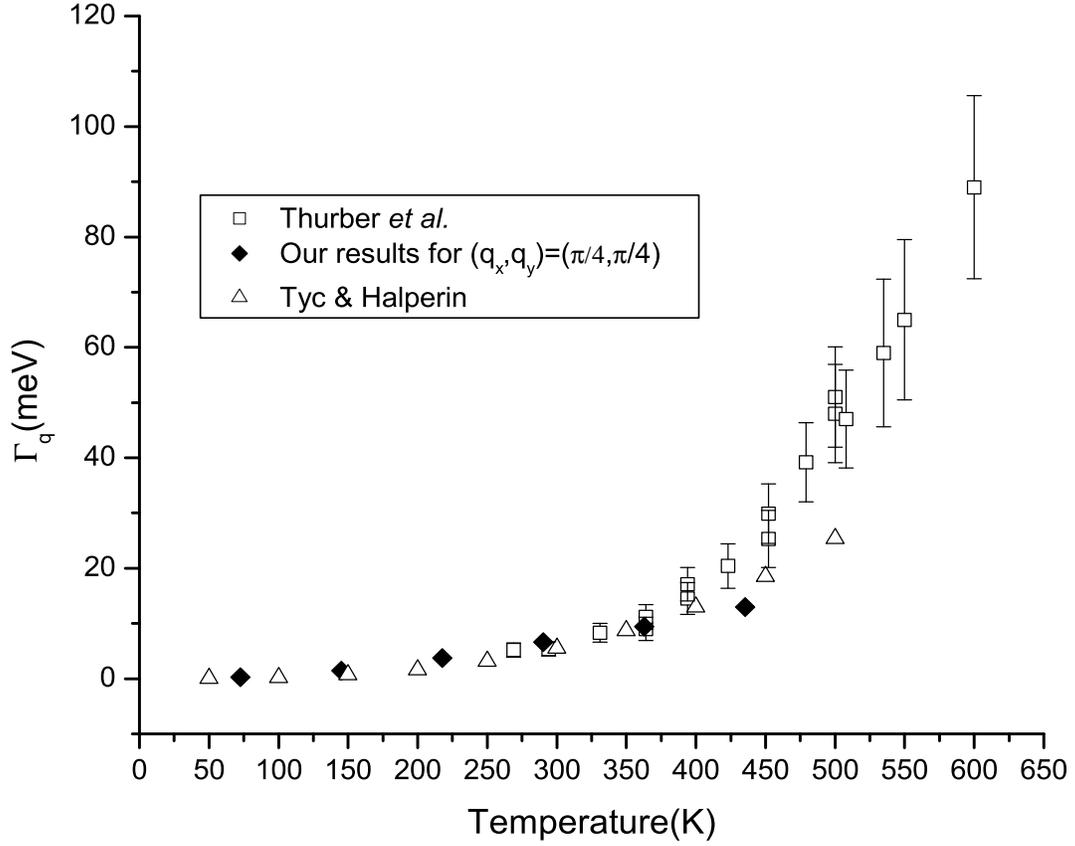}
\end{center}
\caption{
Spin-wave damping, $\Gamma (\vec{q},\tau)$,
as a function of temperature
for $(q_x,q_y)=(\pi/4,\pi/4)$ (filled-diamonds). 
Open squares are the experimental
results from Thurber  {\it{et al.}}, 
\cite{Thurber} and open triangles
are theoretical results from Tyc and Halperin 
\cite{Tyc}.
}
\end{figure}

\end{document}